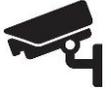

# A review of modern surveillance techniques and their presence in our society

Alexis Roger

University of Montreal, alexis.roger@umontreal.ca

Technology is now omnipresent around us. Especially with the recent health crisis, many people started working remotely, bringing home an additional computer. Combining this with our smartphones that we could never leave behind, we are always surrounded by these technological marvels. However, they come along with a rather dark side from which many people choose to look away, preferring to live in denial: the surveillance. All of these devices can be used to keep a close eye and ear on us. The modern surveillance machine has reached a new, groundbreaking, size; and we will attempt to understand how we ended up in this situation. To have a complete understanding of the problem, it is important to gather some historical background to comprehend where this issue comes from as well as a review of the different actors. Each actor has a specific skillset it will use to acquire the desired information, and what information they choose to gather depends strongly on their motives. We will go over the many tricks used to gather our information, as well as its relevance in the current surveillance climate.

# 1 INTRODUCTION

A recurring theme in our society today is our personal data. Who gathers it? Who controls it? And to whom does it belong? To begin, in developed countries, the population fervently objects any type of intrusion of the government in their personal life, refusing to share any details with it. These very same people then turn around and hand vast amounts of their data to major international corporations. This duality in behavior has led to some very interesting characteristics in our modern society. The divide between what is accepted coming from the government and what is coming from companies has led to there being many gray zones in the rules and regulations, which are readily exploited by the different actors. One of the underlying points in these debates around data acquisition and usage is surveillance.

Surveillance is the act of collecting and analyzing data on an individual in order to gather as much information about them as possible. The gathered information may then be used to find patterns in an individual's routine, to predict his behavior, assist law enforcement and general intelligence gathering. These predictions may then be used for commercial purposes, in order to create targeted advertising, tailored to a citizen or group. These can also be used to identify a weakness in the routine, in order to target an enemy. Furthermore, they can be used to identify any unusual behavior, which could reveal hidden secrets. Surveillance has taken on many forms over the years, but in the last couple decades, with the advent of globalization and of the internet, it has been taken to a whole new scale. Surveillance has crept on all of us for the so-called common good, and we should recognize its presence and spread in order to choose. Choose to let it go as-is; choose to regulate it, the choice falls on all of us.

Surveillance today is nothing like it has been in the past. In this paper we will attempt to cover its genesis, up to its lightning evolution in recent times. We will also review the different techniques used and how they have expanded over time. Along with this we will shed light on the many actors that are at play here, mainly behind the scenes, and from where comes their vested interest in these surveillance techniques. Finally, we will conclude with what may be done to attempt to contain the modern surveillance issue.

# 2 HISTORY OF SURVEILLANCE

## 2.1 From the genesis to the pre-industrial era

In the sixth century before Crist, Darius the Great ruled over the vast riches of the Achaemenid Empire. This empire stretched from the Greek isles in the west all the way to the Indian sub-continent to the east. One of its regional rulers was a tyrant whom went by the name of Histiaeus, and he resided on the Aegean Sea. He was later appointed to govern a town on the Persian Gulf but appreciated it very little and began plotting his return. His plan was simple: he would initiate a revolt on the Aegean Sea and the emperor would have to send him back. However, in order to do this he would need to communicate discreetly with his alies over two thousand kilometers away. To achieve this, he chose to shave the head of his most trusted slave, and tattoo a message on it. Once his hair had regrown, he sent him on the month-long journey across the continent, for him to be re-shaved upon reaching his destination; finally revealing the message. This story is most commonly known as the first recorded use of steganography, transcribed by the Greek historian Herodotus [1]. However, more importantly for us, if Histiaeus resorted to steganography, it meant that he did not wish to simply send a tablet or scroll, which would have been faster than waiting for the hair of the chosen slave to grow back. This therefore goes to show that Histiaeus feared that he was under surveillance and that his communications would be monitored. Therefore, this story is also the first recorded occurrence of surveillance.

We can also observe the intense use of reconnaissance units by many civilizations during the antiquity. This shows the most prominent use case for surveillance techniques at that time: gaining a tactical advantage during a military engagement.

These soldiers, also known as scouts, had the crucial role of scouting ahead of the main contingent of the army, finding adequate paths and surveilling enemy troops. Some anecdotes survive to this day, recounting how the romans would send scouts to surveil a land they were planning on concerning, analyzing both the amount and disposition of the local population, along with geographical and topographical features of the terrain. A notable example of this is the scouting of the African coast by roman ships during the third Punic war [2]. Rome also had some deeply divided political factions and although little firsthand accounts have survived the test of time, it is told that the different factions would strongly surveil each-others movement and meetings, especially inside the city, leading to all the stories of plots and betrayals which have been passed down to us. Cicero would lament that: "I cannot find a faithful message-bearer. How few are they who are able to carry a rather weighty letter without lightening it by reading", to the fellow scholar Atticus [3].

These 3 types of surveillance would continue rather unchanged for the coming millennia. The army kept on using recon units to scout terrain and size up an army, later evolving into spies and infiltrating foreign nations; the political elite would continue to employ people whose jobs were to keep a close eye and ear on any rival or untrusted individual, which have led to the French saying "walls have ears" in the XVIIth century; and finally epistolary surveillance, where ones letters would be intercepted. This last point became a very hard and time-consuming endeavor as wax seals became more prevalent, but remained strong especially with the consolidation of mail into national enterprises.

**2.2 The emergence of radio and surveillance in the first half of the XXth century**

All of this radically changed at the end of the XIXth century with James Clerk Maxwell defining the theoretical frame of radio waves, and later Heinrich Rudolf Hertz demonstrating said concepts in practice. This ground-breaking discovery lead to the very first wireless transmissions, with no exchange of physical media. This fundamentally changed the surveillance game. A radio transmitter and receiver combination can communicate with each-other, however they cannot do so exclusively. Any communication issued by the transmitter would be from one to many. In other words, an unlimited amount of receivers, providing they are in range, could hear the transmission. This changed everything. This form of wireless communication was relatively easy to deploy, compared to telegram wires for instance, as it only required an antenna on each side. It therefore became extremely popular and spread like wildfire. However, the development of the radio brought on a major change regarding surveillance: eavesdropping. It was now possible, with a few well-placed radio-receiver stations, to listening in on any radio exchange in the vicinity. This meant that in some way, radio transmissions where now part of the public domain. Anyone could hear every communications, regardless of its goals and intended recipient. This lead to a fundamental shift in the surveillance game: we were no longer trying to hide a communication, as it was as effective as screaming on a public forum, we were now trying to disguise them, more like screaming in English in a French public forum. This brought on a resurgence of a field which had not seen much recent innovation: cryptography.

Surveillance, in the sense information-gathering, was now easier than ever. However, interpreting the gathered information just became significantly harder. The information wars between different actors complexified with different innovations being debuted during the Great War, also known as World War I. However, in these early days of the technology, many mistakes were made. These mistakes introduced critical vulnerabilities in the systems which they were meant to secure. For instance, in 1917, the United States was surveying the German ambassador to Washington and they intercepted a message, known as the Zimmermann telegram [4] which had been issued by the German Foreign office. This telegram spoke of plans to forge an alliance between Mexico and Germany, in order to discourage the United States from joining the war in Europe. This message was decrypted by the British Admiralty's Room 40, and actually had the opposite effect, leading the United States to ally with the French and English and join the war in Europe. This event showed every modern power how critical it was to have a robust encryption method, and how lucrative it could be to have powerful cryptanalysts in order to run successful and effective surveillance operations. Combined with this, recent years had seen incredible advances in the development of electromechanical devices. This lead Gilbert Vernam to propose, in 1926, the founding ideas of symmetric key encryption schemes [5]. During that time, the airplane also took off and quickly became the first independent aerial surveillance platform.

Following on the lessons learnt during the First World War, the British were keen to reiterate the success of room 40 in the Second World War. They consequently establish an even larger codebreaking and surveillance apparatus, which took on the name of the chateau in which it was located: Bletchley Park. Other nations tried establishing similar surveillance centers, however none of them came close to being as efficient as Bletchley Park. The English had the Germans under very heavy radio surveillance, and it is thanks to all the transmissions that they recorded that Allan Turing was finally able to see cracks in their encryption scheme and to eventually build a machine to exploit them and break the code [6]. This came from the study of many messages, leading to a pattern being spotted: no letter would encode into it-self and regular messages, such as meteorological reports, contained predictable phrases, such as the day, station name and salute. Other nations would specialize more in other surveillance fields. To name the main ones, the United States made considerable advances in the field of radar, which gave the allied powers a particularly effective aerial warning system, by surveilling the skies for enemy aircraft. This was rapidly followed by the sonar, which is the under-water equivalent. This allowed for underwater surveillance, permitting the first steps in submarine-tracking technologies. Although crude at first, these methods rapidly progressed. For their part, the Germans started developing the art of triangulation in order to locate pockets of French resistance. The concept of triangulation is rather simple: by having an array of antennas with known relative location, and seeing which antennas pick-up a given transmission, it is possible to deduce that the emitter is between all of these antennas. The relative distance to each antenna can then be estimated with the relative power of the transmissions received. As the Germans realized, the same result can be achieved using a single antenna moving which known bearing and speed and studying the evolution of the power of the transmission received over time [7]. These techniques are still in use today, with the static version being used to track cellphones thanks to the cell tower grid, while the dynamic version is being used a lot in mountain search and rescue, with the Recco and Arva systems being utilised extensively after avalanches to find survivors. We will study these modern applications later on in this survey. We can note that war really is the catalyst of innovation.

**2.3 The Cold War surveillance revolution**

During the cold war, which opposed the United States and the Soviet Union, many surveillance techniques where developed, and the ground work for many modern technologies was laid down during this time. Both governments surveilled both their opponent and their own population, which lead to many innovations in the field of intelligence gathering being developed.

To begin, the Cold War brought back spies to the center of the surveillance game. Both sides sent agents in the other's territory, or sphere of influence, with a variety of missions. Of course, one of their main mission was intelligence gathering and reporting, but they could have addition roles, such as sabotage or diffusing propaganda and mayhem in a population, or have domains of expertise, such as gathering state, trade or nuclear secrets [8]. Even though both sides engaged in these activities, they had vastly different approaches. The Americans would survey sectors of interest and when they saw an opportunity with one of the members of this sector, they would attempt to bribe or convince them to betray their country to defect. On the other hand, the Russians would survey their opponents and then send one of the infamous Sparrows to conduct sexpionage [9]. This was a type of surveillance and intelligence gathering where the agent would use sexual favors or promises of intimacy to extract information from a target. Berlin, a city divided between the two ideologies, was a spy nest. It was a major diplomatic spot with many important meetings between representatives of the superpowers meeting there. This lead to there being many spies of all nations were based in the city to surveil and report on the activity of the others. Berlin also had a very significant location for the west: it was landlocked in the middle of East Germany, and by extension the eastern bloc. This had important surveillance significance for other surveillance techniques.

Another surveillance method which was as prevalent as ever, was the constant monitoring of all radio frequencies. As the British had already started during the Second World War, every nation now began building dedicated, massive, technical centers whose role was to monitor all of the radio wave spectrum to attempt to intercept and surveil the communications of the other parties. These surveillance efforts could be monitoring anything. For instance, looking at discrepancies between the information shared with the public on news channels and other, less public and official, sources could reveal where the most propaganda was being made, and therefore, where the regime was having issues, and what it was attempting to cover-up. Furthermore, it could, for instance, be used to intercept pilot communications, allowing the opposite side to gain valuable information if they decided to impersonate a pilot for an illegal transport. It would also reveal crucial information about the adversary's military infrastructure: where are they emitting from? Which what power? About what? What frequency or regularity? This may give an idea of what type of military installation was being monitored, and what was its mission. This information would then be used to devise possible attack plans. For this reason bases like this would be scattered all over the globe by both parties.

The Cold War also brought the art of aerial surveillance to a whore new level, literally. At the end of World War 2, both the United States and the Soviet Union got their hands on the jet propulsion technology that was developed by the Germans during the war. This was, at that time, the most advanced jet propulsion program and propelled both nations into the jet age. The most iconic American bomber of the Second World War was the B-17 Flying fortress. This propeller-driven bomber would have a cruising altitude of around 30 thousand feet, about 9 kilometers, and a speed of 150 miles per hour, about 240 kilometers per hour. In sharp contrast with this aircraft, the Americans developed the U-2 high-altitude jet-powered reconnaissance aircraft which flew at double the altitude and at more than three times the speed [10]. This aircraft, which started flying for the Central Intelligence Agency in 1955, was fitted with a camera which had the impressive resolution of 2.5 feet, or about 75 centimeters, from its regular cruising altitude. Everything at that time being analogue, the aircraft would record meters of tape as the pilots flew over the Soviet Union and these tapes would later be analyzed by CIA teams to surveil enemy movements, infrastructure projects, military base activity, or in Cuba the placement of the Russian nuclear missiles. A decade later, the now aging aircraft was replaced with the new Lockheed SR-71 Blackbird [11]. This was basically a highly upgraded version of the previous aircraft, flying faster and higher with even better surveillance equipment, such as an infrared camera. Not much is known about the soviet aerial surveillance programs as they have continued to keep them well-guarded secrets, while the United States has declassified many projects and missions. Nevertheless, considering how neck-in-neck both nations where and the constant spying, it seems reasonable to assume the Soviet Union had their own surveillance aircrafts.

Additionally, using all the experience they acquired in jet propulsion, both blocs, the West and the East started shooting for the stars. The space race was portrayed to the public as a national objective, putting both ideologies directly against one-another, in order to prove which one was stronger. In fact, even though it did serve this objective, it also had the important mission of intelligence gathering. Mere years after the Russians managed the feat of putting a man-made object in orbit, Sputnik 1 in 1957, both the Soviet Union and the United States had spy satellites orbiting the earth and surveilling each-others movements [12]. These satellites where sometimes mounted with cameras. However, due to the technical limitations at the time and there orbits being predictable, spy planes were still preferred to gather precise data on a specific installation. Still, they were able to provide a more macroscopic view of a territory for broader studies. Nevertheless, these satellites, by being always up compared to planes, fulfilled the crucial mission of missile launch detection. From that point on, neither side could fire a missile without the satellite constellation of the other picking it up. For example, the Vela satellite constellation had the goal of identifying any nuclear explosion by monitoring the neurons and gamma rays present in the upper atmosphere.

Each side was now spying on the other with a combination of satellites, planes, monitoring stations and ground personnel. One may ask why? Why was all of this surveillance put in place? Why was an obscene amount of money poured in programs that could keep you informed about what was happening on the other side of the world? The answer to this is the core principle of the Cold War, and why it never became a direct conflict: Mutually Assured Destruction. Both nations needed to keep tabs on what their adversary was doing, as if he became stronger faster, he may be able to deal a decisive blow without any repercussions. And conversely: if one became significantly stronger than one's adversary, it may be possible to attempt to neutralize him while remaining relatively unscathed, showing once and for all the dominance of one's ideology and forcing the rest of the world to adopt it. This lead to the proliferation of surveillance techniques and it is also during this time that the shift from analogue to digital picked up speed, culminating with the birth of the internet in the eighties, just years before the dissolution of the Soviet Union.

**3 SURVEILLANCE IN THE MODERN AGE**

Even though surveillance in the modern age was built on the previously developed surveillance methods, much has been done to upgrade and modernize them. The rapid pace at which technology has evolved, and the vested interests of the different actors in the surveillance space to always stay one step ahead has meant that tremendous efforts have been placed by the many actors to stay on the bleeding edge of innovation. We will now go through the different "levels" of surveillance, starting with the global surveillance apparatuses, then zooming in to national surveillance schemes and finally we will go over the privatization of surveillance. For each of these scales we will study the different methods used by the multiple actors in order to achieve their goal of information control. For each scale, we will then do a case study, taking a singular example to show how it allies different surveillance sources to extract as much information as possible and how it then uses that information, whether it is for monitoring, decision making or profiting.

**3.1 A global surveillance apparatus**

Following the dissolution of the Soviet Union at the end of 1991, one may think that the surveillance efforts by different countries would slow down. However, this was not the case. As we have previously discussed, many surveillance satellites had been placed in orbit at an enormous cost, so they were not just going to be turned off. Pandora's Box had been opened and it was not going to be possible to close it back up. Any country with enough motivation and capital soon began launching their own satellites into space. Joining the United States and Russia, China has launched a constellation of new high-resolution satellites with the latest reported launch late last year, in November 2021 [13]. The United States has also upgraded their own satellite capabilities in the past years [14], and although it is suspected that Russia has followed suit, no conclusive evidence exists.

As discussed before during the Cold War, satellites have their limitations. That is why aerial reconnaissance continues to be a widely used. This is especially true for places with rapidly evolving situations as a satellite cover may not always be possible. However, the means used for this reconnaissance have evolved drastically. Gone with the on-board crew, the United States now operates unmanned aerial vehicles, also known as drones. These drones are remote controlled from a secure base and deployed primarily in zones of conflicts. With their ten hour autonomy and the possibility of pre-programmed flight paths, these drones can pretty much fly themselves, along with their entire armada of surveillance equipment. Being lower than satellites, these drones run a higher risk of being downed. Still, thanks to the resolution and rapid response which they bring, the risk is well worth the reword. Furthermore, as they are ground based, their sensors can be continuously upgraded so that they always carry the most up-to-date and performant version of a sensor. Speaking of sensors, the United States Air Force primary drones, the Predator and Reaper drones, are equipped with a multitude of sensors: a variable aperture TV camera as well as a variable aperture thermographic camera provided a live and controllable view in both day and night conditions respectively [15]. It also possesses a synthetic aperture radar which is unaffected by smoke or bad weather but provides lower resolution images. The first drone, the Predator, was mainly a reconnaissance and surveillance platform. On the other hand, the Reaper drone has built on its predecessor by adding the "attack" specification, leading to it being better armed but nevertheless keeping its surveillance cameras. For obvious security reasons, the exact technical characteristics of the cameras which the drones currently use is not public knowledge. The only limit on these drones is there fuel, as crews can easily be changed out when tired. To solve this issue, the solution of aerial refueling is currently being studied, with the United States Navy publically showing off their progress mid-2021 with a successful drone aerial refueling [16]. It goes without saying that other countries are most likely developing reconnaissance drones, however none do so as publically as the United States. Sparse reports have appeared stating that the systems on board these drones is now so advanced, thanks to advances in artificial intelligence, that no human operator is necessary for simple reconnaissance tasks. These reports also mention that target acquisition is also being more and more automated. Although these reports are few, far between and akin to conspiracy theories, this paints a grim picture of the Reaper. Indeed, seeing the problems that public artificial intelligence is managing to solve for object identification within an image, i.e. YOLO v5 [17]; and how even commercial airlines have highly automated systems, it is not too farfetched to believe that we are entering a dystopian future, where an all-seeing drone may become judge, jury and executioner.

Coming back down to earth, and continuing down below the waves, we will find our next global surveillance technique. The internet as we know it today operates with a giant web of fiber optic cables crisscrossing the ocean floor and transferring information at the speed of light. All the information flowing across the internet passes through these cables, therefore tapping them may reveal a treasure-trove of data. The goal of performing a fiber tap is to gain access to the information shared on a link without interrupting the actual connection, which would alert the users that something is off. This may be done by bending a fiber as far as it will go before breaking and if the radius is tight enough, some of the light will blead out, not being totally reflected inside, and in effect duplicate the signal. This is in fact noticeable as there would be a slight increase in the attenuation of the signal. However, as all underwater cables above 400 kilometers have repeaters which boost the signal, a well-placed tap could be unnoticeable. Articles also mention that tapping the repeaters themselves is the easiest way to perform a tap as all the fibers are laid out neatly instead of being bundled together in the cable [18]. Having worked with a major submarine cable producer, they believe that the security of the information which transit through the cables is the responsibility of the user. Furthermore, these repeaters can easily be thousands of meters deep, encased in many layers of protective housing for all the sensitive equipment to remain intact on the sea floor. All of this natural protection makes taping a modern deep-water fiber link, such as one between the United-States west coast and Asia, extremely challenging and expensive. Nevertheless, I would not underestimate a superpower and do believe it is still a possibility to tap certain, very targeted, cables. These would be cables inking military infrastructures or untrusted nations.

Finally, there are vast international wireless monitoring efforts which take place and we will study the Five Eyes alliance to understand it. It should be noted that other alliances exist but this is the most proficient and well-known one. The Five Eyes alliance is an alliance between the five most powerful English speaking countries across the globe in an effort to share information and to cover each-others blind spots [19]. This alliance regroups the United Kingdom in Europe, the United States and Canada on the American continent, and Australia and New-Zealand on the other side of the pacific. Following the declaration of the "war on terror" by the United States, this alliance started heavily monitoring the digital space. The Five Eyes alliance was described as a "supra-national intelligence organization that does not answer to the known laws of its own countries" by Edward Snowden, a former National Security Agency contractor. To illustrate this, he provided documents which showed that while no participating agency could directly monitor its own citizens, they would all monitor each-others and then share the information. This lead to precisely the situation which was meant to be illegal and prohibited in the first place, with the intelligence agencies being, in effect, surveilling their own citizens. The initial system was based primarily on phone and fax networks, by tapping public switched telephone networks and satellite communications. However, since the attacks of the 11$^{th}$ of September 2001, the apparatus has been upgraded to tap public internet infrastructure, such as the fiber optics network mentioned above as an upstream collection method. A second approach is also used in order to have access to a wider range of data: the PRISM surveillance program [20].

The PRISM surveillance program is a program launched by the United States National Security Agency following the 2001 terrorist attacks. This program collects internet traffic handled by different internet companies, such as Google LLC and Yahoo. The data that these companies need to turn over, upon request, is "any data that match court-approved search terms", as defined under Section 702 of the FISA Amendments Act of 2008. Additionally, this disposition allows the National Security Agency to request companies turn over data that may have been encrypted while it was traveling over the internet, saving the hassle of tying to decrypt it. Nevertheless, the National Security Agency remains very proficient at code breaking. The National Security Agency is also the one that defines the cryptographic standards that are implemented and used by the United States government, and more generally the internet as a whole. For instance, it was them who propagated the Advanced Encryption Standard, the first publically available high encryption scheme. It was latter shown to have vulnerabilities with longer keys being weaker than they appear, although it remain a robust encryption schemes [21]. The real issue came with the Dual Elliptic Curve Deterministic Random Bit Generator which was a cryptographically secure pseudorandom number generator. It was shown that by knowing two hidden constants, which described the elliptic curves, it was possible to decrypt any message that had been encrypted using the dual elliptic curve scheme [22]. The NSA having developed the algorithm, they were immediately accused of having purposefully added the backdoor but the claims were dropped due to lack of proof. With the revelations of Edward Snowden, some internal documents seemed to suggest that this was indeed a plot by the NSA to undermine the internet security of their adversaries, allowing them to easily harvest the information.

To conclude, the Five Eyes, led by the American NSA, conducts a thorough global surveillance program which aims at "protecting the United States of all enemies, foreign and domestic". The efficiency of this program cannot be accurately measured by independent sources due to its very secretive nature. Most of the information we have regarding this program stem from leaks, mainly the one performed by the whistleblower Edward Snowden when he realized that he, an American, was forced to spy on his fellow countrymen. His objection came from the fact that he believed they were operating against the constitution, namely the fourth amendment which protects Americans from unreasonable searches and seizures. This means that we cannot cross-check the information provided with any other primary source, which have lead some people to question the validity of these documents portraying the NSA as the all-knowing one. Regardless, this amount of surveillance and power cannot be entrusted to a single entity, as the dual elliptic curve backdoor showed. In order to continue having a safe and free internet, free from surveillance and discrimination, we must attempt to challenge these forces when possible, requesting more transparency by them and more control by their government.

**3.2 Focusing on national surveillance schemes**

To change the focus a bit, we have seen how the United States proceeds in running an international surveillance apparatus, surveilling everyone across all borders. We will now change our focus and see how a government which wishes to surveil their own population without any legal constrains may go about doing so. We will go over both physical and digital techniques that may be used together or in a stand-alone fashion.

The obvious place to begin with this is with the cameras, and more specifically the closed circuit surveillance cameras. These cameras have been popping up in cities all around the globe in the last two decades under the claim of "improving public safety". Original systems were expensive, bulky and required a professional installation and maintenance crew. But as the technology improved, the cameras progressively shrank in size, and became more independent with lighter infrastructures. They also became cheaper to purchase. This big reduction in cost to purchase and operate is one of the driving factors behind their proliferation. Cameras were no longer something only high-end security firms would have but something that could be found in every mall and neighborhood stores. Seeing the incredible asset these camera systems could be for policing, many cities also started installing cameras across the city's public areas for the police. Some countries managed to regulate them, for instance in France a public camera may not show an apartment's door or window; they can also not be monitored in real time. If the French police want to access the cameras, they can only do so when investigating a specific crime with a warrent. Nonetheless, Paris has a camera density of 255 cameras per square kilometers. We could compare this to other cities such as Singapore, a city-state with an unfounded negative view from Europeans. Singapore may be qualified as a city-state which maintains a high oversight on its' population, but it barely has 120 per square kilometer. The highest count in the western world goes to London, with about 400 per square kilometer, and where the cameras may be monitored live. This camera network is one of the most powerful tools readily accessible to law enforcement for subject tracking. This overseeing eye sees everything in the city and are starting to make some people a little bit uneasy. However, this is the "cost of safety", and far too many are still willing to make that tradeoff. All camera figures come from a Surfshark investigation [23].

Other than simply tracking individuals on the side-walk, there are many other methods that can be used to track a subjects' movements more generally: To begin, in every city with a public transport network, regular users would have their subscription directly on a transport card, assigned to them. This allows for the efficient tracking of intra-city movements and is why, during the Hong Kong protests of 2019, protesters would buy single use tickets with cash [24]. This surveillance is particularly prevalent in European cities, with robust public transportation networks. This is most likely one of the main downsides to public transportation: the ability to efficiently analyze a populations movements. Furthermore, inter-city transit is handled by four types of travel: by car, by bus, by train and by plane. Let us put the plane aside for now. Starting with the car, more and more countries are installing toll gates. These toll gates may be on the highway, as in France, or at the city entrance, as in London. Regardless, this infrastructure can give a general idea of where an individual of interest may be and maybe even clue to where he is moving. For instance, if two week-ends in a row you drive from Paris to Bordeaux, it is fair to assume that the third weekend, as you are leaving Paris, you are headed to Bordeaux. By bus or train may be the most anonymous forms of travel as a simple name is required upon purchasing the tickets. However, many of us use discount passes that are often referenced with unique serial numbers, leading once again to a tracking possibility. Finally we have the plane. Since the 11th of September 2001, the security measures surrounding air travel have been completely reformed and airports have become some of the most surveilled places. Combine with this the multiple identity checks and you have the most controlled and trackable form of transportation between two cities, especially if they are international. The most extreme case of this airport surveillance is the Tel Aviv Ben Gurion airport [25]. This airport combines social engineering, rigorous background searches and an extensive CCTV network in order to monitor all passengers closely, in addition to monitoring the flights around the airport. Some qualify their amount of security and racial profiling overboard, however the airport is a completely safe haven with many dangers surrounding it. In this very specific case, I do believe that the outstanding amount of surveillance is justified.

Something that has been highly discussed recently by many countries is digital currency. Following the trend of more and more of our lives moving online, and most of our payments following along, many countries are currently debating creating a digital equivalent of their currency. This has also been triggered by a governments seeing the attractiveness of cryptocurrencies such as Bitcoin and Ethereum, and fearing that their population would migrate regardless. And of course, if people want to shift to a different currency style, they want to keep some sort of control. Still, a digital currency and a cryptocurrency are not to be confused. The main difference is that a digital currency is issued by the central back, hence by extension the government [26]. On the other hand, cryptocurrencies have no overseeing regulatory body. As it stands today, the banks are the ones able to surveil their users, tracking them by using their credit and debit card usage data; however, with a digital currency, the government would be able to view every transaction and hence surveil the spending habits of its entire population. It has also been theorized that with a digital currency and smart contracts, a government would be able to lock certain spending habits and hence change its population's usage of their money, to achieve a greater goal, such as boosting certain sectors of the economy. This government control would help securing the currency, making counterfeits, embezzlement and tax avoidance harder, but at the expense of sacrificing part of the population's control and freedom, bringing on a new level of institutionalized surveillance. As of today, China it the country the most advanced in the development in a digital currency, with the digital Yuan being in the works since 2014. Many trials have been run in different cities to iron-out the persisting issues and get people used to the idea. However, there is still no clear plan on when a national rollout will take place. India is hot on their heels trying to institute a digital equivalent to the rupee, which they hope to roll out in the coming years [27].

In the age of the internet, in which we are now, we have voluntarily introduced one of the most powerful surveillance tools into our daily lives: the smartphone. For now we will focus mainly on the cellular side of it and we will revisit it along with the rest of its surveillance capabilities later on. Cellphone providers have tremendous amounts of data on us. They know to who belongs every number, and therefore any communication between two parties is known. They also have the ability to locate users using previously discussed triangulation methods. In most places, these service carriers are also the internet service providers for the homes and businesses of their users. This therefore leads them to having access to massive amount of data: they know with who you communicate and which internet resources you use, as in which websites you visit and what servers you regularly exchange with. Even though most of the traffic is encrypted this has two major limitations: some services or rural areas are still operating on unsafe networks, such as HTTP pages, which transmit data as clear text, or the 2G mobile network [28]. Using these unsecure protocols allows the carrier to view precisely what you are exchanging about with the website and the contents of your discussion. To counteract this, HTTPS was introduced and has become whether well adopted. It uses transport layer security (TLS), with a secure sockets layer (SSL) certificate, to encrypt traffic to and from a secured server. The internet service provider is now unable to see the contents of all the communications passing through its network. However, does that really matter? Stewart Baker, former NSA general counsel said the following: "Metadata absolutely tells you everything about somebody's life. If you have enough metadata you don't really need content". As we have seen before, the NSA is notoriously good at surveillance. The idea behind this quote is that it is possible to surveil an individual, even if he is using encrypted communications. This is because all information traversing the internet does so in packets, and each packet has a certain amount of public information so that the network nodes know how to handle it. For instance, the source and destination address along with the protocol is part of this public metadata. Server addresses for major services are known and in the domain name system, the address repository of the internet, is a public ledger. For instance, it is well known that google.com has a domain name server at the address 8.8.8.8. A single packet is not useful as it provides very little information. However, considering all of our traffic is routed through them, our internet service providers see it all: what websites, at what frequency and with which habits we visit. For instance: a user requesting an umontreal.ca website, may it be stadium or mail, on a daily basis may be a student or professor at University of Montreal. Looking at his other traffic, such as if he connects to videogame servers, would allow us to choose which one. On the other hand, if an individual connects to umontreal.ca, mcgill.ca during the month of December, this is most likely a prospective student looking at the different programs and deciding where to apply. If they also consider polymtl.ca, they may be more interested in doing engineering and in French, while if they also visit Concordia.ca they would be more interested in pursuing studies in English. This is how and internet service provider, or a national organization monitoring all local traffic, may be able to surveil the individuals on the network using exclusively metadata. The server-side applications of this type of surveillance and ways to protect against it will be seen later on, in the part about corporate surveillance.

Following the techniques laid down above, it is possible for a nation to easily monitor its population. In spite of this, it remains rather hard to control the flow of information, and what they can access. We will now look at China, and how it goes above and beyond the methods studied above to surveil, track and control its population [29].

To begin, the Chinese population operates within the confines of the Chinese Great Digital Wall. This wall is active at every internet access point of the country and blocks any unauthorized traffic from entering. As an example of this, many United States based applications and social media, such as Facebook, are blocked within China. Even though the contents are encrypted, surveillance techniques based on the metadata of the request, as discussed above, allow this blocking to take place. Now that the Chinese population cannot have access to foreign services, which Chinese government does not control, it started working on domestic equivalents. As they control what the masses use, the Chinese government has created an "app for everything", known as WeChat. The vision of this app is that you never need to leave it, so that, regardless of what phone you purchase, you install it and you have access to everything and the government has all of its surveillance tools installed by the same occasion. WeChat covers everything from social media and messaging to online payments and food deliveries. All the information of all the users, what they look at, what they research, where they are, is then freely shared with the Chinese communist party as part of a mass surveillance network. As it is not possible to use anything else, as all other platforms are blocked, and combined with the network effect, where if someone you wish to exchange with uses an app you will use the same, WeChat has an uncontested monopoly in China, even if downloading it is equivalent to opting in to the Chinese surveillance machine.

Furthermore, this digital surveillance is far from the only surveillance technique used by the Chinese regime to monitor its population. Since 2009, the Chinese government has been experimenting with a Social Credit System [30]. This system works by giving each citizen a score and when they contribute positively to society, they win points and when they break a rule they loose some. To illustrate this: volunteering in one's neighborhood would grant points while jaywalking would remove some. The amount of points a person has possesses a tremendous impact on their life. The access to healthcare, education and the possibility to travel is all guided by the amount of social credit one has. Lower scores leads to lower quality of life with lower healthcare quality and higher interest loans. The gamification of the social credit coupled with the lack of the ability to protest in China has actually lead to its adoption going rather smoothly, and is now deployed nation-wide. The scoring used here may actually be manipulated in order to weed out undesirable behaviors, leading to an easier population to control. Yet, in order to implement such a system, it was necessary that the Chinese communist party deploy the greatest surveillance apparatus ever. To this end, the government has invested in CCTV cameras, with Chinese cities holding the 14 positions out of the top 15 cities with the most CCTV cameras per 1000 people, being rivaled only by London. The highest of them all is Taiyuan which boasts an impressive 120 cameras per 1000 people, equating to 1 camera for every 8 individuals [23]. It would be infeasible to review this extraordinary amount of footage by hand, therefore the Chinese government stays on the bleeding edge of artificial intelligence technology, identifying individuals with their biometric features, mainly facial recognition. Artificial intelligences then try to make sense of the images captured and if they detect an illegal behavior, such as jaywalking, they will automatically deduct points on the individuals social credit score.

The amount of surveillance in China, both in the physical world with the cameras and virtually thanks to the apps and internet censoring, is simply staggering. It has reached a point where countries, such as the United States have recommended that athletes participating the 2022 winter Olympics there should go with only a burner phone, which they would use to download the Chinese apps and destroy it when they return home. China is at the forefront of the surveillance movement and, regrettably, other countries, such as India, seem to be following suit instead of going in the opposite direction. The main risk is that as other governments start to take note how easy it is to control the population in a surveillance state, they may very well follow suit, grinding down the population's resistance little by little.

**3.3 Corporate surveillance in the land of the free**

Many private actors have been attempting to gather as much information about their public as possible. Compared to why governments do it, may it be for a war on terror or manipulating the public's behavior, corporations do it in an effort to maximize their profits. If all started with the mundane membership card: people would sign up to it to receive news and use it at checkout to get points or discounts. By tracking a user's purchase, the store could try to find the users preference and serve them with targeted advertising, hoping that would entice them to visit the store more regularly and spend more. The most famous example of this is the Target pregnancy prediction [31].

In a small town in the United States, a man barges in the local Target, demanding to speak to the manager. The man was furious as his teenage girl, a member at the Target in question, had started receiving ads for baby products. He requesting to know why Target was inciting his daughter to get pregnant and demanded they stop immediately. A few weeks later he called back the manager, apologizing for his previous outburst, as his daughter had confessed indeed being pregnant. Target had figured out by surveilling the girl's shopping habits that she was pregnant before her own father even new about it. This was possible as target had been monitoring women all over their store network to notice subtle changes in purchasing habits which would indicate a coming baby, detected by the purchasing of diapers, baby formula and so one at a later time. Target would try mixing the baby promotional content in with their standard promotional content, yet seems that in this case it was not enough. This story brought to light the corporate surveillance that was quickly becoming the norm. However, this did not hinder the advancements of private surveillance infrastructures, with each field having its own surveillance actors and techniques.

Nowadays, most of the surveillance is done on the World Wide Web. Many different techniques can be used in this online jungle to track and surveil people without their knowledge or explicit consent. One of the simplest is cross-site sign-in. This is when a user authenticate himself for a service or website using a third party service, such as using google.com sign-in to non-Google website or service. This allows the third party, the one providing the authentication method, to tack the user across not only its own services but every other service he connects to using these credentials. This could help the service provider identify a user's behaviors and preferences, which would then allow it to serve them more targeted advertising. As in the case of Target, corporations track their users in order to serve them with more personalized advertisements, which have a higher chance of being viewed by the user, and hence selected. Most services on the internet operate in this fashion, where it is free for the user but companies pay a premium to only advertise to a targeted audience.

This has led to the famous quote by Andrew Lewis: "If you're not paying for the product then you are the product". A good example of this system in action would be Gmail; Google would passively monitor your email conversations and would serve you ads based on them. For example, a discussion about a summer retreat in the Caribbean islands with a friend would lead to advertisements for cheap flights from your home city to different Caribbean islands. This was an experiment that we could all replicate before 2017, when Google said it stopped [32]. While Gmail scanned our messages, it apparently respected our privacy as only automated algorithms ran on our emails, with no employee accessing them. Does this make you feel better?

Another very popular tracking technique which has been in the news for the past few years is cookie tracking. A cookie is a small bit of data that a website leaves on your computer. This cookie can be used to remember your login, but also to contain a unique identifier. The unique identifier can then be followed across many websites. Thanks to all the negative press this method of surveillance has received in the past few years, it is now slowly being set aside, with modern browsers attempting to block it and making it easier for users to remove said cookies. In spite of that, there remain many techniques to track a specific user across the internet. One of the most basic consists in creating a unique browser fingerprint based on the web request that is made. Web requests, especially those requesting JavaScript scripts, contain much information about the user in order to tailor the experience to every device. For instance, using the website amiunique.org, we see that using information about our operating system, browser version, preferred language and time, our request can have a unique web fingerprint. Furthermore, there exists a unique tracking technique, which is one of my personal favorite, thanks to its uniqueness and simplicity: favicon tracking [33]. This method, released in 2021, works regardless of if the cookies and history are cleared and even if the user goes incognito, or in private browsing depending on your browser. The concept is rather simple: the little icons in the browser header bar are known as favicons. These tiny images are cached by the browser in order to limit web traffic, but some researches realized they could exploit this fact: by redirecting users through different subdomains with different favicons, they could avoid the browser redirect warning and create a unique identifier for the user based on which favicon already had and which he requested. This could take a mere second to precisely identify the user and no one is the wiser. This attack worked on all major browsers, with the notable exception of Firefox. It was later revealed that this was due to a bug in the implementation of favicons in Firefox, as closing the browser cleared them. Hence, this is now a feature of Firefox.

The next point that we will explore are VPNs. Known as virtual private networks, these companies sell a promise of anonymous browsing. The core principle is to route all a user's traffic through a central server, so that the server whose webpage or service you are requesting does not know to whom it is talking to and anyone surveilling a server of interest with metadata analysis techniques as seen before is unable to track the incoming traffic. It also offers the same protection on the client's side, obfuscating their communications from their internet service provider. In spite of this, virtual private networks are not the golden shield their advertisements would lead you to believe. In fact, NordVPN and all other providers replace your internet service provider as your single point of access to the web. They can therefore use all the same surveillance tricks as seen before, except now it becomes a for profit enterprise and not, as are many internet service providers, a state regulated company. These VPNs can conduct an extensive and unregulated surveillance of their users by basing themselves in strategic countries. There has been a lot of advertising around VPN services in the recent years, mainly by influential tech reviewers, such as Linus Tech Tips, which are highly misleading. To be clear, VPNs are a crucial part of the internet. May it be to remotely connect to offsite servers or to avoid censorship in countries such as Iran or China, they are necessary; but they are not designed for everyone. Additionally, if everyone uses a VPN, all the internet traffic will be routed through more servers and the big increase in bandwidth requirements may bottleneck the internet a whole. With a large enough view of the network, these also become useless, and the metadata analysis may resume [34]. Allow me to illustrate: if user A request site C by passing through a VPN provided server B, you will have this series of communication: A to B, B to C with the request, and C to B then B to A with the response. Therefore packets transiting through B can be matched when they come in and when they leave by comparing delays and packet size. This may render a VPN useless, and just add additional latency when attempting to connect to a website. Some decentralized alternatives exist, being maintained by generous benefactors and the open source community. This allows these VPNs to not integrate surveillance software. A great example of this is the infamous TOR network, the onion ring. TOR comes with its own issues. For instance, all TOR nodes are public and could therefore theoretically be monitored. Although possible, this requires an immense surveillance apparatus to pull off and is only feasible on a national scale. Other techniques are being studied to try and isolate TOR traffic, may it be for surveillance purposes or to out-right ban it [35]. With the democratization of blockchain, other, new, solutions are starting to appear. For instance, the Mysterium network is a decentralized VPN where users are invited to "lend" their spare bandwidth to allow others to connect through them [37]. This network has the ability of disguising all of the traffic across many smaller nodes and making it blend in to the normal residential traffic. This makes surveillance efforts and traffic isolation considerably harder. On top of this, the network a lot more resilient to attacks, and standard VPN-countermeasures, such as server blacklisting, will not work on it. The Mysterium project is a very nice usage of blockchain, solving a real-world problem, and not just creating a speculative bubble. Replacing standard VPNs with Mysterium would be a step in the right direction, but also brings on its own issues: every node can monitor itself, so any individual on the network can monitor part of it. However this is already a considerable improvement to a single entity surveilling everything that is going on.

A very effective tool for surveillance that companies are currently developing are virtual assistants. Microsoft is developing Cortana, Amazon is working on Alexa and Apple is improving Siri, which are meant to compete with each-other as well as with Google assistant. These assistants are embedded into our devices, in fact it is not possible to setup a Microsoft or Apple product without getting the assistant as a package. These assistants are linked with our calendars, know our likes and dislikes and probably even know us better than we know ourselves. They are the ultimate surveillance device, always by our side, always listening, and we let them in our homes with open arms [38]. Complementing these assistants, there exists an armada of Smart Home gadgets. These gadgets surveil every aspect of our life, going from the temperature to the vacuuming robot creating a floor-plan of your house. But worse of all are the cameras. Both inside and outside Smart Home cameras very often lack the most basic security measures and there have been many problems over the years with cameras that the owners installed to protect themselves being turned into tools used to surveil them. A small sample of what this type of surveillance may look like can be found on this website, which gathers a list of all camera which are open to the internet: insecam.org.

The final point which we will study before focusing on how a singular company exploits all of the previously seen techniques to enable mass surveillance is the global positional system (GPS). The GPS is a service developed by the United States army and later opened to the public. This was a major leap in positioning technology. Anyone with a receiver was now able to know their precise location across the world. This also allowed for an important tracking ability be devices containing a GPS chip. Therefore, any hardware manufacturer can incorporate this chip, sell it to the customer as a feature an then us it to track the device, and by extension the very same consumer. This tracking technology has been so well studied that major technological firms are even able to deduce the means of transport a person is using, and when they are entering and exiting a vehicle. In cities GPS systems, meaning the American GPS, Russian GLONASS and European GALILEO, have difficulties giving a precise location due to the building distorting the radio waves. Therefore, to improve its accuracy, many devices now use a combination of Wi-Fi and Bluetooth to accurately place the device between know markers. To accurately locate the markers, Google uses its Google Street View cars to additionally gather tremendous amounts of network data, such as Wi-Fi networks names, active Bluetooth devices and so on. In and of itself Google Street View is an incredible surveillance apparatus, with the images it gathers being used for all type of demographic studies [39]. The information gathered is then used to triangulate a precise location, based on the initial GPS guess. This GPS information has also been used against the very entity which developed it: in early 2018, the Strava fitness tracking app was being used by United States soldiers deployed in Iraq. The app recorded their runs and would therefore revealed the locations of the bases along with their perimeter wall, patrol routes and other key features [40]. This was a major security issue for the Army and was promptly banned, but not before leaking all of this highly valuable information. The Apple recently released Airtags also bring along a whole new position tracking method: piggy backing nearby phones for connection [41]. It is still to soon to tell what exactly they will bring but I would approach them with a healthy dose of caution.

To follow up with this, technological giants, such as Facebook, now Meta, Amazon, Apple and Google all have their own surveillance systems, however we will focus on one that has had much time in the papers recently regarding it surveillance apparatus: Meta. Meta is an international technological conglomerate worth a whopping 560 billion US dollars at the time of writing. The company holds many regularly used application, such as Facebook, WhatsApp, Instagram and so on; going above 3.6 billion monthly active users. This puts the company in a unique situation to be able to collect enormous amounts of data. Using Facebook friends, it is possible to generate social graphs; using WhatsApp and Messenger you can further detail these graphs by weighting the edges with communications. Natural language processing ran on the messages sent between two people can say a lot on their relationship, i.e. father-son, employer-employee, good friends and so on. Instagram brings this a step further as follower data can be used to establish the social status of and individual or organization within its community. And this is only looking as "simple" social graphs. Using the information gathered around what someone posts, on any platform, from where and with whom (identified either by tags, account geo-positioning or facial recognition), it is possible to create an ever more detailed portrait, containing even more information. Meta is able to track us: where we have been, with who, and how we get along since the very moment we set foot on the platform. This extreme surveillance allows them to serve us with ever more precise advertising. Even though many have tried blowing the whistle on these predatory practices, Meta has the market so well cornered with the network effect and very deep pockets that we keep on coming for more. The network effect is the reason why we join a specific social network, and why we later chose to leave it: if all of the people around us use a specific network, we are more likely to go and join them. In the same way, if the group migrates to a new platform for any reason, there are high chances that we will follow. During the writing if this article, Meta's surveillance hegemony is being more intensely challenged. Apple has recently announced that the surveillance software Meta uses will now be opt-in on its' devices, instead of the previously used opt-out method. Turns out people are not fan of being constantly surveilled, so very little have reactivated the feature. This has led to a drop on the profiling abilities of Meta, leading their stock to plummet by 30% in a day. The surveillance they operate is purely profit driven. They watch us to find our desires, our weaknesses, and then give us what we want. This will make us happy and come back. Meta then charges for tailored advertising, and you understand why they are so popular, and why every time someone attempted to regulate their advertising they would threaten leaving [42].

## 4 CONCLUSION

To conclude, we are now more surveilled than ever. In every aspect of our life and by every device around us. Sometimes it is by foreign governments attempting to acquire information, sometimes it is by our own government evaluating if we pose a threat, and sometimes it is by a corporation trying to leverage our information to turn a profit. It falls on all of us to be aware of these surveillance measures in order to either choose to live with them, either stand up against the current system.

We are currently exchanging our liberty and freedom for more surveillance and security. Left unchecked, this may have drastic consequences. It has been shown time and time again that we act differently when we are under surveillance. Edward Snowden summed-up this effect really well: "under observation, we act less free, which means we effectively are less free". This is known as the Hawthorne effect.

In order to avoid all the aforementioned surveillance issues, with the different parties watching us, many have attempted to develop open-source alternatives that may even be self-hosted. However, without the financial support these solutions have trouble developing. Additionally, it is infeasible to imagine everyone is willing to buy and host a server, spending the time and money learning, setting up and maintaining their own personal infrastructure. Nevertheless, a balance can hopefully be found. Maybe it would be worth starting to pay for services on the internet that respect our right to privacy. Regrettably, it is now engrained so deeply in society that on the internet everything is free that changing the model would be extremely challenging. We are also trading our data for convenience in many cases, and our laziness may very well be our downfall. Especially with the recent issues with people refusing to adopt the Covid-19 vaccination passes from fear of being tracked. An explanation for this behavior is that they are ready to "purchase" a service with their data, which we must now see as a currency, and as they were not getting a service they wanted out of the Covid-19 pass system, they did not want to "purchase" and use it. Covid-19 has not created a new surveillance problem, it has simply sped up the development of an already existing one. Many "protection" measures have been put in place during the crisis, in many if not all countries, such as South Korea. These measures could more accurately be described as "surveillance" measures in most cases. The chances of the governments relinquishing all of this new found power are slim to none. Sadly I fear the current situation is becoming our new normal, and once that happens, new "protection" measures will arrive.

If we wish to remedy this situation and correct our course, we need to act now. We need to start electing an executive and legislative power who understands the right to privacy and who will actively fight to pass laws to protect it. On a more personal level we can also review the amount of information we post online. Do we really need to share a picture of us to the entire world? Maybe to our friends it would be enough. We need to start teaching good habits to children, the most vulnerable: how to recognize a scam, how to protect their information. And, for the most motivated, we can migrate to more user-conscious services, such as decentralized VPNs like Mysterium or encrypted messaging such as Signal. Of course, for real, hardcore, technology enthusiasts nothing beats a properly secure home server, but as discussed this may be a little overboard, especially if it is not properly secured, making it more vulnerable to attacks.

Finally, some may argue that this report on modern surveillance does not concern them as they have nothing to hide, and to those people I leave this very powerful quote from Edward Snowden:

"Arguing that you don't care about the right to privacy because you have nothing to hide is no different than saying you don't care about free speech because you have nothing to say."

**REFERENCES**

[1] Herodotus. -430. The Histories, Book 5; 2003 translation by Aubrey De Selincourt
[2] N. B. Rankov, N. J. E. Austin. 1995. Exploratio: Military and Political Intelligence in the Roman World from the Second Punic War to the Battle of Adrianople
[3] Cicero. 61. Letters to Atticus
[4] Thomas Boghardt. 2003. The Zimmermann Telegram: Diplomacy, Intelligence and The American Entry into World War I
[5] Gilbert Vernam. 1926. Cipher Printing Telegraph Systems For Secret Wire and Radio Telegraphic Communications
[6] Andrew Hodges. 1983. Alan Turing: The Enigma
[7] Maj. Gregory C. Clark. 1997. Deflating British radar myths of World War II
[8] Jennifer Llewellyn, Steve Thompson. 2020. Cold War espionage
[9] Michael Dobbs. 1987. Sexpionage why we can't resist those KGB sirens

[10] U-2 aircraft. https://www.lockheedmartin.com/en-us/products/u2-dragon-lady.html

[11] SR-71 aircraft. https://www.lockheedmartin.com/en-us/news/features/history/blackbird.html

[12] P. Norris. 2009. The Political Impact of Spy Satellites

[13] Andrew Jones. 2021. China launches new Gaofen-11 high resolution spy satellite to match U.S. capabilities

[14] Mike Wall. 2020. US launches 4 secret spy satellites to orbit

[15] 2021. Predator RQ-1 / MQ-1 / MQ-9 Reaper UAV

[16] Megan Eckstein. 2021. US Navy, Boeing conduct first-ever aerial refueling with unmanned tanker

[17] Xingkui Zhu, Shuchang Lyu, Xu Wang, Qi Zhao. 2021. TPH-YOLOv5: Improved YOLOv5 Based on Transformer Prediction Head for Object Detection on Drone-captured Scenarios

[18] Zafar M. Iqbal, H. Fathallah, Nezih Belhadj. 2011. Optical fiber tapping: Methods and precautions

[19] James Cox. 2012. Canada and the Five Eyes Intelligence Community

[20] Ewen Macaskill and Gabriel Dance. 2013. NSA files decoded

[21] Alex Biryukov, Orr Dunkelman, Nathan Keller, Dmitry Khovratovich, Adi Shamir. 2010. Key Recovery Attacks of Practical Complexity on AES Variants With Up To 10 Rounds

[22] Dan Shumow, Niels Ferguson. 2007. On the Possibility of a Back Door in the NIST SP800-90 Dual Ec Prng

[23] Surveillance cities. 2021. https://surfshark.com/surveillance-cities

[24] Ramona Pringle. 2019. Hong Kong protesters go offline to dodge China's digital surveillance

[25] Oren Liebermann. 2016. In airport security, many say Ben Gurion in Israel is the safest

[26] Itai Agur, Anil Ari, Giovanni Dell'Ariccia. 2022. Designing central bank digital currencies

[27] Annabelle Liang. 2022. India says it will launch digital rupee as soon as this year

[28] Mohsen Toorani, A. Beheshti. 2008. Solutions to the GSM Security Weaknesses

[29] Xiao Qiang. 2021. Chinese Digital Authoritarianism and Its Global Impact

[30] Dean Curran, Alan Smart. 2020. Data-driven governance, smart urbanism and risk-class inequalities: Security and social credit in China

[31] Kashmir Hill. 2012. How Target Figured Out A Teen Girl Was Pregnant Before Her Father Did

[32] Selena Larson. 2017. Google will no longer read your emails to tailor ads

[33] Konstantinos Solomos, John Kristoff, Chris Kanich, Jason Polakis. 2021. Tales of Favicons and Caches: Persistent Tracking in Modern Browsers

[34] Chris Stokel-Walker. 2021. VPN flaw could put users at risk

[35] Application of Deep Learning on the Characterization of Tor

[36] Clayton Johnson, Bishal Khadka, Ethan Ruiz, James Halladay, Tenzin Doleck, and Ram Basnet. 2021.Traffic using Time based Features

[37] 2017. Mysterium Network Project Whitepaper

[38] Anouk Mols, Yijing Wang, Jason Pridmore. 2021. Household intelligent personal assistants in the Netherlands: Exploring privacy concerns around surveillance, security, and platforms

[39] Victoria A. Sytsma, Nathan Connealy, Eric L. Piza. 2020. Environmental Predictors of a Drug Offender Crime Script: A Systematic Social Observation of Google Street View Images and CCTV Footage

[40] Alex Hern. 2018. Fitness tracking app Strava gives away location of secret US army bases

[41] Kiyohiko Hattori, Toshinori Kagawa. 2021. Parasite Sensing 2 -- Study on the possibility of measuring human flow using the Apple AirTag

[42] Jillian Deutsch, Stephanie Bodoni. 2022. Meta Renews Warning to EU It Will Be Forced to Pull Facebook